\documentclass[sigconf,screen,nonacm]{acmart}

\usepackage{booktabs}
\usepackage{tabularx}
\usepackage{enumitem}
\usepackage{xcolor}
\usepackage{tikz}
\usepackage{graphicx}
\usepackage{fontawesome5}
\usepackage{balance}

\usepackage[framemethod=default]{mdframed}
\definecolor{coreargbg}{RGB}{255,255,255}
\definecolor{coreargline}{RGB}{0,0,0} 
\newmdenv[
  backgroundcolor=coreargbg,
  linecolor=coreargline,
  linewidth=1.0pt,
  roundcorner=3pt,
  innertopmargin=6pt,
  innerbottommargin=6pt,
  innerleftmargin=8pt,
  innerrightmargin=8pt,
  skipabove=6pt,
  skipbelow=10pt
]{coreargbox}

\usetikzlibrary{arrows.meta,fit,positioning,backgrounds,calc}

\setcopyright{none}
\renewcommand\footnotetextcopyrightpermission[1]{}

\title{Trusting Trust in AI Software Engineering Agents: \\ The Why, What, and How?}

\author{Pranav V}
\affiliation{
  \department{RISHA Lab}
  \institution{Indian Institute of Technology}
  \city{Tirupati}
  \country{India}
}
\email{cs25s523@iittp.ac.in}

\author{Sridhar Chimalakonda}
\affiliation{
  \department{RISHA Lab}
  \institution{Indian Institute of Technology}
  \city{Tirupati}
  \country{India}
}
\email{ch@iittp.ac.in}

\author{Vibhu Saujanya Sharma}
\affiliation{
  \institution{Accenture Labs}
  \city{Bangalore}
  \state{Karnataka}
  \country{India}
}
\email{vibhu.sharma@accenture.com}

\author{Vikrant Kaulgud}
\affiliation{
  \institution{Accenture Labs}
  \city{Bangalore}
  \state{Karnataka}
  \country{India}
}
\email{vikrant.kaulgud@accenture.com}

\begin{abstract}
AI software engineering agents are increasingly integrated into development workflows, from autocomplete assistants to autonomous repair agents and multi-agent software development pipelines. Trust matters because these agents can produce, revise, and pass on artifacts that developers, other agents, or organisations may later rely on under uncertainty and risk. Yet trust in such systems is often treated as a single measurable property of an AI system. This is insufficient for software engineering, where trust depends on who relies on the agent, how that reliance is structured, and which properties must hold in that deployment.

We propose a structural framework for reasoning about \emph{trust} in AI software engineering agents. Inspired by the recursive nature of trust in software systems, we decompose trust into three analytical questions: \textbf{Why trust?}, identifying the trust scenario; \textbf{What trust?}, identifying the trust configuration; and \textbf{How much trust?}, identifying the  trust properties. From these questions, we derive three scenarios, four configurations,
and five trust-relevant properties, state which of these are closed
by construction and which remain open to revision, and instantiate
the resulting profile on a ChatDev-style multi-agent workflow. The
framework is conceptual, and is offered as a foundation for the
empirical work that would validate it, rather than as a numerical
trust metric.
\end{abstract}


\begin{document}

\maketitle

\section{Introduction}

Trust in software is commonly understood as a relationship
between a trustor and a trustee. The trustor decides to rely on
the trustee under uncertainty and risk~\cite{lee2004trust,mayer1995trust}. Thompson's
Reflections on Trusting Trust~\cite{thompson1984trusting} pointed to a different way in
which trust can arise. A compiled binary may behave correctly,
but the user who runs it is also relying, implicitly, on the
chain of compilers and tools that produced it. Trust in the
binary therefore depends on artifacts and processes that the user
cannot directly inspect.
\begin{coreargbox}
\noindent\textbf{Core Argument:}
Trust in AI software engineering agents is not
only about a single agent's output. When agent outputs become inputs
to other agents, or persist as artifacts that downstream users depend
on, the trustworthiness of the final artifact may depend on the chain
that produced it, the concern Thompson raised for compilers, now
arising at the scale of multi-agent pipelines. Trust models centred on
a single trustor-trustee pair do not directly address this.

We propose a structural framework that decomposes trust into three
questions: \textbf{WHY trust?} identifies the trust scenario;
\textbf{WHAT trust?} identifies the trust configuration; \textbf{HOW
MUCH trust?} identifies the trust properties. Benchmarks such as
SWE-bench and HumanEval primarily evaluate one configuration and do
not capture trust concerns that arise in iterative or propagated
settings. The contribution is not a final trust metric, but a
question-based framework for analysing trust in AI-agent software
engineering workflows through scenarios, configurations, and
trust-relevant properties.
\end{coreargbox}

AI software engineering agents appear to introduce a similar pattern.
Modern agents may modify code, invoke tools, interact with repositories,
coordinate with other agents, and produce software artifacts that persist.
Multi-agent software development frameworks make this pattern more
concrete. For example, ChatDev proposes a chat-powered software
development framework in which specialised LLM-based agents collaborate
through a structured chat chain across design, coding, testing, and
documentation phases~\cite{qian2024chatdev}. In such a workflow, the final
software artifact is not produced by a single isolated agent. Product goals
may shape requirements, requirements may shape architecture, architecture
may shape code, and code may shape review, testing, and later
revisions~\cite{qian2024chatdev}.

Thus, the trust question is no longer only whether a developer should trust
one agent's output. It is also whether one agent should rely on another
agent's intermediate output. This pattern is not directly addressed by
classical trust models, which treat trust as a relationship between two
parties rather than a property that can flow through artifacts. We argue that
this artifact-flow pattern is a fundamental trust concern raised by agentic
software engineering, and that existing trust models do not fully address it.

Existing work on trustworthy AI and AI agents identifies useful dimensions
such as transparency, accountability, reliability, robustness, fairness,
alignment, autonomy, and governance~\cite{aleti2026trustworthy,cheng2026threepillar,hong2024iclr,wagner2024ieee}.
These perspectives are valuable because they identify properties and controls
that trustworthy AI systems should possess. However, for SE agents, trust also depends on the deployment structure:
who or what relies on the agent, how reliance is configured, and whether
agent-produced outputs are accepted, reused, or passed downstream~\cite{mayer1995trust,marsh1994formalising,lee2004trust}.
This is especially visible in multi-agent software development workflows, where
one agent's intermediate output may become another agent's input or part of the
final artifact~\cite{qian2024chatdev,aleti2026trustworthy}.


\begin{figure*}[t]
\centering
\resizebox{0.96\textwidth}{!}{%
\begin{tikzpicture}[
    font=\footnotesize,
    every node/.style={align=center},
    >=Stealth,
    line cap=round,
    line join=round,
    agentbox/.style={
        draw=#1, rounded corners=5pt,
        fill=#1!8,
        minimum height=1.10cm, minimum width=3.5cm,
        align=center, inner sep=5pt, line width=0.7pt
    },
    widebox/.style={
        draw=#1, rounded corners=6pt,
        fill=#1!7,
        minimum height=1.15cm, minimum width=4.8cm,
        align=center, inner sep=5pt, line width=0.7pt
    },
    legendbox/.style={
        draw=black!30, rounded corners=5pt,
        fill=white, inner sep=10pt,
        align=left, line width=0.7pt
    },
    deleg/.style={
        -{Stealth[length=2.2mm]}, thick,
        draw=delegcol, shorten >=2pt, shorten <=2pt
    },
    assist/.style={
        -{Stealth[length=2.2mm]}, thick,
        draw=assistcol, shorten >=2pt, shorten <=2pt
    },
    proc/.style={
        -{Stealth[length=2.2mm]}, thick,
        draw=proccol, shorten >=2pt, shorten <=2pt
    },
    prop/.style={
        -{Stealth[length=2.2mm]}, thick,
        draw=propcol, shorten >=2pt, shorten <=2pt
    }
]

\definecolor{chatgptcol}{RGB}{16,163,127}
\definecolor{claudecol} {RGB}{160,65,135}
\definecolor{ccodecol}  {RGB}{50,90,190}
\definecolor{copilotcol}{RGB}{205,120,20}
\definecolor{aidercol}  {RGB}{120,70,170}
\definecolor{sweagcol}  {RGB}{35,145,160}
\definecolor{loopcol}   {RGB}{145,75,150}
\definecolor{finalcol}  {RGB}{55,150,70}
\definecolor{orgcol}    {RGB}{50,80,170}

\definecolor{assistcol}{RGB}{39,174,96}
\definecolor{proccol}  {RGB}{142,68,173}
\definecolor{delegcol} {RGB}{41,128,185}
\definecolor{propcol}  {RGB}{231,76,60}


\node[font=\bfseries\small] at (0,7.05)
    {Trust Configurations in a Multi-Agent Software Development Workflow};
\node[font=\itshape\scriptsize, text=black!60] at (0,6.70)
    {A virtual software company for software development};

\node[widebox=orgcol, minimum width=5.2cm] (user) at (0,5.55) {};
\node[font=\bfseries\small] at ($(user.center)+(0, 0.18)$) {User / Developer};
\node[font=\scriptsize]     at ($(user.center)+(0,-0.18)$) {software task};

\coordinate (delegLabelPos) at ($(user.west)+(-1.4,0.2)$);
\node[anchor=east, font=\bfseries\small, text=delegcol]
    at (delegLabelPos) {Delegated Trust};
\node[anchor=east, font=\scriptsize, text=black!55]
    at ($(delegLabelPos)+(0,-0.38)$) {user delegates a task};


\node[agentbox=chatgptcol] (ceo) at (-5.0,3.75) {};
\node at ($(ceo.west)+(0.42,0)$)
    {\includegraphics[height=0.90cm]{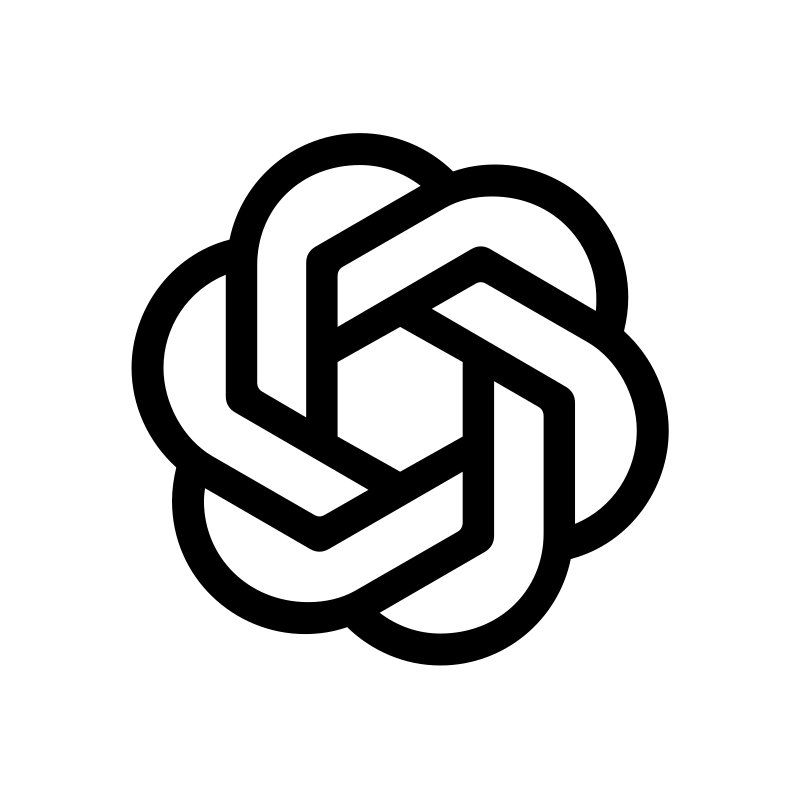}};
\node[font=\bfseries\small] at ($(ceo.center)+(0.24, 0.15)$) {CEO Agent};
\node[font=\scriptsize]     at ($(ceo.center)+(0.24,-0.18)$) {product vision};

\node[agentbox=claudecol] (cpo) at (0,3.75) {};
\node at ($(cpo.west)+(0.42,0)$)
    {\includegraphics[height=0.75cm]{logos/anthropic.png}};
\node[font=\bfseries\small] at ($(cpo.center)+(0.24, 0.15)$) {CPO Agent};
\node[font=\scriptsize]     at ($(cpo.center)+(0.24,-0.18)$) {requirements};

\node[agentbox=ccodecol] (cto) at (5.0,3.75) {};
\node at ($(cto.west)+(0.42,0)$)
    {\includegraphics[height=0.75cm]{logos/anthropic.png}};
\node[font=\bfseries\small] at ($(cto.center)+(0.24, 0.15)$) {CTO Agent};
\node[font=\scriptsize]     at ($(cto.center)+(0.24,-0.18)$) {architecture};

\node[widebox=copilotcol, minimum width=4.6cm] (prog) at (0,1.85) {};
\node at ($(prog.west)+(0.48,0)$)
    {\includegraphics[height=0.90cm]{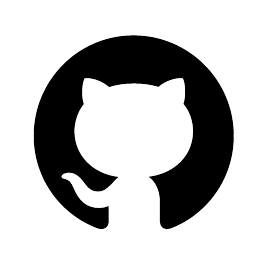}};
\node[font=\bfseries\small] at ($(prog.center)+(0.32, 0.15)$) {Programmer Agent};
\node[font=\scriptsize]     at ($(prog.center)+(0.32,-0.18)$) {implements code};


\node[agentbox=aidercol] (rev) at (-3.5,-0.15) {};
\node at ($(rev.west)+(0.42,0)$)
    {\includegraphics[height=0.75cm]{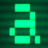}};
\node[font=\bfseries\small] at ($(rev.center)+(0.24, 0.15)$) {Reviewer Agent};
\node[font=\scriptsize]     at ($(rev.center)+(0.24,-0.18)$) {code review};

\node[agentbox=sweagcol] (test) at (3.5,-0.15) {};
\node at ($(test.west)+(0.42,0)$)
    {\includegraphics[height=0.75cm]{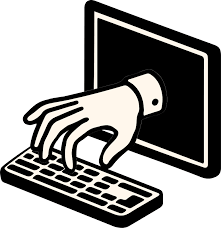}};
\node[font=\bfseries\small] at ($(test.center)+(0.24, 0.15)$) {Tester Agent};
\node[font=\scriptsize]     at ($(test.center)+(0.24,-0.18)$) {tests};

\node[widebox=loopcol, minimum width=5.2cm] (loop) at (0,-1.90) {};
\node[font=\large, text=loopcol] at ($(loop.west)+(0.48,0)$) {\faSync};
\node[font=\bfseries\small] at ($(loop.center)+(0.32, 0.15)$) {Iterative Improvement};
\node[font=\scriptsize]     at ($(loop.center)+(0.32,-0.18)$) {feedback and fixes};

\node[widebox=finalcol, minimum width=4.8cm] (final) at (0,-3.55) {};
\node[font=\large, text=finalcol] at ($(final.west)+(0.48,0)$) {\faBox};
\node[font=\bfseries\small] at ($(final.center)+(0.32, 0.15)$) {Final Software Artifact};
\node[font=\scriptsize]     at ($(final.center)+(0.32,-0.18)$) {code, tests, documentation};

\node[legendbox, minimum width=4.2cm, minimum height=4.0cm] (legend) at (9.8,1.55) {};
\node[font=\bfseries\small, anchor=north] at ($(legend.north)+(0,-0.30)$)
    {Trust Configurations};

\draw[assist] ($(legend.west)+(0.42, 0.90)$) -- ++(0.95,0);
\node[anchor=west, font=\small, text=assistcol]
    at ($(legend.west)+(1.55,0.90)$) {\textbf{Assistive Trust}};

\draw[proc]   ($(legend.west)+(0.42, 0.18)$) -- ++(0.95,0);
\node[anchor=west, font=\small, text=proccol]
    at ($(legend.west)+(1.55,0.18)$) {\textbf{Process Trust}};

\draw[deleg]  ($(legend.west)+(0.42,-0.55)$) -- ++(0.95,0);
\node[anchor=west, font=\small, text=delegcol]
    at ($(legend.west)+(1.55,-0.55)$) {\textbf{Delegated Trust}};

\draw[prop]   ($(legend.west)+(0.42,-1.28)$) -- ++(0.95,0);
\node[anchor=west, font=\small, text=propcol]
    at ($(legend.west)+(1.55,-1.28)$) {\textbf{Propagated Trust}};


\coordinate (bus) at (0,4.95);
\draw[draw=delegcol, thick, line cap=round, shorten <=2pt]
    (user.south) -- (bus);
\draw[draw=delegcol, thick, line cap=round]
    (-5.0,4.85) -- (5.0,4.85);
\draw[deleg] (-5.0,4.85) -- ([yshift=2pt]ceo.north);
\draw[deleg] (bus)        -- ([yshift=2pt]cpo.north);
\draw[deleg]  (5.0,4.85) -- ([yshift=2pt]cto.north);

\draw[assist]
    ($(user.east)+(0,0.10)$) -- ++(2.7,0) -- ($(cto.north)+(0.3,0)$);
\node[anchor=west, align=center] at ($(user.east)+(2.75,0.52)$)
    {{\bfseries\small\color{assistcol}Assistive Trust}\\[1pt]%
     {\scriptsize\color{black!55}developer guidance}};

\draw[proc] (ceo.east) -- (cpo.west);
\draw[proc] (cto.west) -- (cpo.east);

\draw[prop] (ceo.south) |- ($(prog.west)+(0,0.15)$);
\path (ceo.south) -- (-5.0,1)
    node[midway, left=6pt, align=center]
        {{\bfseries\small\color{propcol}Propagated Trust}\\[1pt]%
         {\scriptsize\color{black!55}vision informs code}};

\draw[prop] (cpo.south) -- (prog.north);

\draw[prop] (cto.south) |- ($(prog.east)+(0,0.15)$);

\draw[prop] ([xshift=-2pt]prog.west) -| ([yshift=2pt]rev.north);

\draw[prop] ([xshift=2pt]prog.east)  -| ([yshift=2pt]test.north);


\draw[proc] ($(test.north)+(-0.2,0)$) -- ++(0,0.40)
            |- ($(prog.east)+(0,-0.15)$);
\node[right=6pt, align=center] at (1,0.8)
    {{\bfseries\small\color{proccol}Process Trust}\\[1pt]%
     {\scriptsize\color{black!55}review/test feedback}};

\draw[proc] ([yshift=-1pt]rev.south) |- ([xshift=-3pt]loop.west);

\draw[proc] ([yshift=-1pt]test.south) |- ([xshift=3pt]loop.east);

\draw[prop] (loop.south) -- (final.north);
\path (loop.south) -- (final.north);

\end{tikzpicture}%
}
\caption{Trust configurations in a multi-agent software development workflow. The structure is inspired
by ChatDev~\cite{qian2024chatdev} and populated with contemporary commercial agents (ChatGPT, Claude, Claude Code,
GitHub Copilot) and open-source agents (Aider, SWE-agent). Role bindings are illustrative; the same agents
can occupy other roles in practice. The framework is independent of specific tool-role bindings. See Section~4.5 for an explicit mapping of this workflow onto the
framework's scenarios, configurations, and properties.}
\label{fig:chatdev-worked-example}
\end{figure*}

\section{Motivating Example}


Before deriving the framework, we ground its motivation in a concrete example. Figure~\ref{fig:chatdev-worked-example} shows a multi-agent software development workflow where a developer coordinates contemporary AI agents to produce software. The workflow is inspired by ChatDev~\cite{qian2024chatdev} but populated with named commercial and open-source agents to reflect the current agentic SE landscape. ChatGPT contributes vision and planning, Claude elaborates requirements, Claude Code produces architecture and design, GitHub Copilot implements the code, Aider performs review, and SWE-agent runs testing and repair. \textit{The workflow illustrates a recurring pattern: trust is not located in any single agent. It is distributed across the chain, and the trustworthiness of the final artifact depends on the chain that produced it.}


Three trust scenarios are visible directly in this workflow. The user assigning and reviewing the task represents Developer-Agent trust. The reliance among ChatGPT, Claude, Claude Code, Copilot, Aider, and SWE-agent represents Agent-Agent trust, because each agent may depend on outputs produced by other agents. If the final code, tests, or documentation are deployed in an enterprise repository, the case also raises Organisation-System trust concerns.

Four trust configurations are present concurrently. The user delegates the software
task to the agent workflow, creating Delegated Trust. The repeated
planning, reviewing, testing, and fixing interactions create Process Trust. The movement from requirements to architecture, code,
review, tests, and final artifact creates Propagated Trust, because one
agent's output becomes another agent's input. Assistive Trust appears
where the developer guides, checks, or constrains the workflow.

Different properties become salient in different parts of the workflow. In Propagated Trust (Claude or Claude Code outputs feeding Copilot), Contextual Groundedness, Verifiability, and Failure Impact are  most critical because early mistakes propagate into code and tests. In Process Trust (Copilot-Aider-SWE-agent iteration), Functional Adequacy and Stability matter because each iteration should improve the artifact without introducing inconsistent revisions.

Trust models typically assume one party relies on another. In this workflow, multiple parties rely on each other simultaneously. When Copilot acts on Claude Code's architecture, Copilot relies on Claude Code. But the developer, who never re-reads the architecture line by line after approving it, also relies transitively on Claude Code through Copilot's implementation. The trustor-trustee pair is not unique: it depends on which artifact and which agent one is asking about. Empirical work on AI-assisted development confirms that trust is dynamic rather than fixed, shifting as agent autonomy and artifact persistence change ~\cite{sabouri2025icse, liang2024icse}. This motivates a structural decomposition of trust, which we develop in Section 4.
\label{sec:motivation}
\section{Background and Related Work}
\subsection{Trust, Automation, and Trustworthy AI}

The relationship between trustor and trustee has been formalised in
several ways. Mayer, Davis, and Schoorman describe trust as
willingness to be vulnerable to another party's actions based on
expectations about that
party~\cite{mayer1995trust}.
Marsh extends trust to computational settings by showing how
artificial agents can make trust-based decisions under
uncertainty~\cite{marsh1994formalising}. Lee and See emphasise
appropriate reliance in automation, where both over-trust and
under-trust are problematic~\cite{lee2004trust}. These accounts
support our view that trust in SE agents depends on the relying
actor, the conditions of reliance, and the consequences of failure.



Recent trustworthy AI and AI-agent frameworks identify important dimensions such as transparency, accountability, reliability, robustness, fairness, and alignment~\cite{aleti2026trustworthy,cheng2026threepillar}. Empirical work has begun mapping how these dimensions manifest in developer-AI interaction: A survey of 410 developers found that initial suggestion acceptance rates remain low even among daily users, suggesting that trust calibration is itself an active developer practice~\cite{liang2024icse}. Sabouri et al. and team further show that trust in AI-assisted development is dynamic, evolving with experience and context~\cite{sabouri2025icse}. These findings leave room for more explicit reasoning about how trust changes when SE agents modify repositories, coordinate with other agents, or produce artifacts that persist downstream.

Arora et al.~\cite{arora2026delegated} approach agentic AI from a requirements engineering perspective, proposing a "delegated-autonomy boundary", what may be delegated, under what graduated authority, with what oversight, before deployment. Their focus is on design-time specification. On the other hand, Aleti et al.~\cite{aleti2026trustworthy} provide a set of properties that trustworthy SE agents should exhibit, several of which overlap with our own. What this paper adds is the deployment structure that
determines which of those properties matter in a given setting. A
property list on its own does not indicate whether Functional
Adequacy or Failure Impact should take priority for a specific
deployment; the scenario and configuration in which an agent operates
determines that priority. Our contribution is not a competing
property set, but the structure that selects among an existing one.



\subsection{Need for Structural Trust Reasoning}

Software engineering introduces specific trust challenges because
agent outputs often become persistent artifacts~\cite{qian2024chatdev}.
A generated patch, test case, or pull request may be reused,
merged, or passed to other agents, so trust is not limited to
immediate interaction quality but also concerns how outputs affect
later artifacts and workflows. By \textit{structural trust reasoning},
we mean reasoning that connects trust evaluation to the deployment
structure: the trust relationship, the form of reliance, and the
properties that matter in that setting, rather than treating
low-autonomy assistance and high-autonomy repository modification
as equivalent~\cite{chen2021humaneval,jimenez2024swebench}.

Two cases make the point concrete. Consider first a single generated
patch that fixes a failing test. In an Assistive configuration, the
developer reads the patch before accepting it, so Functional Adequacy
and Contextual Groundedness carry the weight: the patch must be correct
and must fit the surrounding code, and the developer will catch it if it
does not. Place the identical patch in a Delegated configuration, where
the agent commits it without review, and the salient properties change.
Correctness still matters, but Failure Impact now dominates, because
nobody will notice a subtly wrong patch until it fails in production,
and Stability matters because the same agent will make the next such
decision unsupervised. The artifact has not changed; the structure of
reliance has, and with it the evidence a practitioner should demand.

The second case concerns an architecture specification. Within a Process
configuration, where the CTO and CPO agents refine requirements and
design across several exchanges, Stability is what matters: the design
should converge rather than oscillate as the agents iterate. Once that
same specification is handed to the implementing agent, the
configuration becomes Propagated, and Contextual Groundedness becomes
the salient property, because an assumption that was harmless while the
design was still being discussed becomes load-bearing once code is
written against it. Failure Impact rises with it, since the assumption
is now embedded in artifacts that later agents will read as settled.

A property list alone cannot express either shift, because it has no
place to record which configuration the output sits in. This is what we
mean by structural trust reasoning: the deployment structure is not
context surrounding the trust question but part of the question itself.

\section{Framework Derivation and Components}

We derive the framework by asking three analytical questions about
trust in AI software engineering agents: \textit{Why trust?}, \textit{What trust?}, and \textit{How much trust?} These questions correspond to three components of the framework: scenarios, configurations, and properties. The framework is primarily descriptive, characterising where trust is situated and
what would need to hold for it to be warranted; this descriptive
account can in turn support evaluation and design, though we make no
claim that it performs all three equally.

The three sets that follow are not derived to the same degree. The
three scenarios are a closed enumeration over the level at which the
trustor is situated: an individual developer, an AI agent, or an
organisation. Each level aligns with a tradition already surveyed in
Section~3 --- automation trust for the individual
operator~\cite{lee2004trust}, computational trust for
agents~\cite{marsh1994formalising}, and governance-oriented trust for
the organisation~\cite{wagner2024ieee}. The four configurations follow from two independent axes. An
autonomy axis (Assistive vs.\ Delegated) samples the two ends of the
autonomy continuum described by Feng et al.~\cite{feng2025levels},
minimal versus substantial human control before an output takes
effect, rather than the finer levels in between. A persistence
axis (Process vs.\ Propagated) reflects the distinction at the centre
of Thompson's concern~\cite{thompson1984trusting}: whether an artifact
outlives the interaction that produced it, or does not. Two binary
axes yield four configurations in total. The five properties, by
contrast, are not closed by construction. They are a consolidation of
recurring concerns from the trust and trustworthy AI literature and
from SE-specific agent failure modes~\cite{aleti2026trustworthy}. This set is not claimed to be exhaustive, and
later work may revise it.

\subsection{Why Trust?}

The first question asks why a trust relationship arises in a given
deployment: because someone, or something, relies on an agent's
output under uncertainty. Answering it requires identifying the
trustor and the trustee.
This follows prior accounts of trust that view it not as a fixed property of a system, but as a context-dependent judgement about whether relying on another actor is warranted~\cite{mayer1995trust,marsh1994formalising,lee2004trust}.
We identify three scenarios.

\textbf{Developer--Agent} trust occurs when an individual developer
relies on an AI agent for SE tasks such as code completion, debugging,
documentation, or localised code changes. The developer remains the
primary decision-maker and may inspect, modify, or reject the agent's
output~\cite{lee2004trust}.

\textbf{Agent--Agent} trust occurs when one AI agent relies on another
agent's output. This matters in multi-agent workflows because assumptions
or errors from one agent may become input to another agent without direct
human inspection~\cite{qian2024chatdev,marsh1994formalising, hong2024iclr}. Both ChatDev and MetaGPT demonstrate that intermediate-agent outputs become inputs for downstream agents with limited semantic validation between stages.

\textbf{Organisation--System} trust occurs when a team, organisation,
repository, or software system relies on an agent's behaviour or outputs.
Here, trust concerns include governance, accountability, risk, and
long-term artifact impact~\cite{hong2024iclr, wagner2024ieee}.

These scenarios are not mutually exclusive. A single deployment may involve
multiple scenarios. For example, a multi-agent development pipeline may
involve Agent--Agent trust between agents while also raising
Organisation--System trust concerns when the resulting artifacts affect a
shared repository.

\subsection{What Trust?}

The second question asks how trust is structured. We identify
four configurations along two dimensions: agent autonomy
(Assistive vs.\ Delegated) and artifact persistence
(Process vs.\ Propagated).
Along the autonomy axis, \textbf{Assistive Trust} describes
low-autonomy settings where the human remains the main
decision-maker~\cite{lee2004trust,vaithilingam2022chi}, while
\textbf{Delegated Trust} describes settings where the agent is
given responsibility to modify code, run tools, or open pull
requests, requiring stronger oversight~\cite{wagner2024ieee}.
Along the persistence axis, \textbf{Process Trust} covers
iterative SE activities such as generation, testing, and
revision, where trust depends on reliability across repeated
steps~\cite{lee2004trust}; \textbf{Propagated Trust} covers
settings where agent-produced artifacts outlive the original
interaction and are reused by developers, tools, or other
agents~\cite{thompson1984trusting}.
Configurations can combine with different scenarios. For example, Assistive
Trust is common in Developer--Agent interactions, Delegated Trust is central
when agents act on repositories, and Propagated Trust becomes important when
agent-produced artifacts are reused by teams or other agents.

\subsection{How Much Trust?}

The third question asks which properties must hold, and on what
evidence, for trust to be justified in a given scenario and
configuration --- not a quantity of trust, but the grounds on which it
rests.


\textbf{Functional Adequacy} refers to whether the agent's output satisfies
the SE task, requirement, or intended functionality. This property
consolidates concerns related to correctness and technical
quality~\cite{aleti2026trustworthy}.

\textbf{Contextual Groundedness} refers to whether the agent's behaviour is
grounded in the project context, repository, APIs, constraints, and tools.
It captures concerns such as hallucinated APIs, unsupported dependencies,
and incorrect codebase assumptions~\cite{aleti2026trustworthy}.

\textbf{Verifiability} refers to whether the agent's output, assumptions,
and changes can be checked, tested, or explained. It supports debugging,
accountability, and trust calibration~\cite{cheng2026threepillar}.

\textbf{Stability} refers to whether the agent behaves consistently across
similar prompts, environments, or code changes. It captures concerns related
to reproducibility and robustness under small variations~\cite{aleti2026trustworthy}.

\textbf{Failure Impact} refers to how severe the agent's failures are, how
easily they can be detected, and how easily systems can recover from them.
It is especially important for delegated and propagated settings, where
agent actions may have downstream consequences through the chain of artifacts
they produce~\cite{thompson1984trusting,qian2024chatdev}.

Table~\ref{tab:propertymap} shows how the trust concerns identified in
prior SE-agent work map onto these five properties. Most fold directly
or in part; three are excluded because they concern the agent's conduct
or efficiency in general rather than whether a specific act of reliance
is warranted. Stability has no direct antecedent in either source set
and is drawn from the broader robustness literature~\cite{aleti2026trustworthy},
which suggests that consistency across repeated runs is less examined in
current SE-agent trust discussions than correctness or transparency.


\begin{table*}[t]
\centering
\caption{Mapping from trust concerns in prior SE-agent work to the five
properties. Partial indicates the source concern is only in part captured
by the property. Excluded entries carry a stated reason.}
\label{tab:propertymap}
\small
\begin{tabular}{p{3.8cm}p{3.6cm}p{3.0cm}p{4.6cm}}
\toprule
\textbf{Source concern} & \textbf{Source} & \textbf{Property} & \textbf{Note} \\
\midrule
Technical quality & Aleti et al.~\cite{aleti2026trustworthy} & Functional Adequacy & Direct \\
Correctness & Roychoudhury et al.~\cite{roychoudhury2026agentic} & Functional Adequacy & Direct \\
Transparency and accountability & Aleti et al.~\cite{aleti2026trustworthy} & Verifiability & Direct \\
Explainability and transparency & Roychoudhury et al.~\cite{roychoudhury2026agentic} & Verifiability & Direct \\
Epistemic humility & Aleti et al.~\cite{aleti2026trustworthy} & Verifiability & Partial: an agent that signals the limits of what it knows supports checking, but humility is broader than checkability \\
Compliance with conventions & Roychoudhury et al.~\cite{roychoudhury2026agentic} & Contextual Groundedness & Direct \\
Team practice alignment & Roychoudhury et al.~\cite{roychoudhury2026agentic} & Contextual Groundedness & Partial: captures fit to project conventions; the social dimension is not a property of the output \\
Security & Roychoudhury et al.~\cite{roychoudhury2026agentic} & Failure Impact & Partial: severity of a security failure is in scope; prevention is a control, not a property \\
Maintainability & Roychoudhury et al.~\cite{roychoudhury2026agentic} & Failure Impact & Partial: maintainable output is easier to repair once a failure is detected \\
Performance & Roychoudhury et al.~\cite{roychoudhury2026agentic} & Excluded & Efficiency does not bear on whether reliance is warranted \\
Societal and ethical alignment & Aleti et al.~\cite{aleti2026trustworthy} & Excluded & Concerns agent conduct in general rather than a specific reliance relationship \\
Bias and ethics & Roychoudhury et al.~\cite{roychoudhury2026agentic} & Excluded & As above \\
Collaboration & Roychoudhury et al.~\cite{roychoudhury2026agentic} & Excluded & Captured by the Process configuration rather than by a property \\
\bottomrule
\end{tabular}
\end{table*}

\subsection{Contextual Trust Profile}

A deployment can be characterised as a contextual trust profile:
\[
\textit{Trust Profile} =
\langle \textit{Scenario (3)}, \textit{Configuration (4)}, \textit{Properties (5)} \rangle
\]
It records where trust is situated, how reliance is configured, and which properties need evidence in that setting.

The trust profile is not a single fixed characterisation.
By combining scenarios, configurations, and properties,
it generates a space of profiles that can be used to
reason about trust in different deployment situations. A
developer using autocomplete, an organisation deploying an
autonomous repair agent, and a multi-agent pipeline each
instantiate different profiles, and each requires different
evidence to justify trust.

\subsection{A Worked Instantiation}
\label{sec:worked-instantiation}

Section~2 describes the workflow; here we instantiate trust profiles for
eight of its reliance relationships, including a transitive
relationship not drawn as edge in
Figure~\ref{fig:chatdev-worked-example}. The table is intended to
demonstrate the profile space rather than to enumerate every
relationship: it covers all four configurations and all three scenarios,
and is representative of the kind of profile a practitioner could build
for their own pipeline. 

Three combinations do not appear. Assistive Trust
presupposes a human trustor and is therefore available only in the
Developer--Agent scenario, which excludes Agent--Agent/Assistive and
Organisation--System/Assistive. Organisation--System/Process does not
arise here because the organisation in this workflow relies on outcomes
rather than participating in the iteration loop.

\begin{table*}[t]
\centering
\caption{Trust profiles for reliance relationships in the workflow of
Figure~\ref{fig:chatdev-worked-example}. Row 6 is transitive: the developer never inspects the architecture directly, but relies on it through Copilot's implementation. The table demonstrates
the profile space rather than enumerating every relationship in the
workflow.}
\label{tab:profile}
\small
\begin{tabular}{p{1.5cm}p{1.7cm}p{1.9cm}p{1.9cm}p{1.5cm}p{2.2cm}p{3.2cm}}
\toprule
\textbf{Trustor} & \textbf{Trustee} & \textbf{Artifact} & \textbf{Scenario} & \textbf{Config.} & \textbf{Salient Properties} & \textbf{Evidence} \\
\midrule
Developer & Agent workflow & task specification & Developer--Agent & Delegated & Failure Impact, Verifiability & audit log of actions taken without review \\
Developer & Claude Code & architecture under guidance & Developer--Agent & Assistive & Functional Adequacy, Contextual Groundedness & the developer's stated constraint is visible in the delivered architecture \\
Claude Code & Copilot & architecture spec & Agent--Agent & Propagated & Contextual Groundedness, Verifiability & implementation traceable to the spec it followed \\
Copilot & Reviewer, Tester & code revisions & Agent--Agent & Process & Functional Adequacy, Stability & test pass rate and diff size across iterations \\
Copilot & SWE-agent & test execution & Agent--Agent & Delegated & Stability, Failure Impact & test outcomes reproducible across repeated runs \\
Developer & Claude Code (via Copilot) & implemented architecture & Developer--Agent & Propagated & Verifiability, Failure Impact & a code-level fault traceable to the architectural decision that caused it \\
Organisation & Agent workflow & merged changes & Organisation--System & Delegated & Failure Impact, Stability & rollback path for any change the pipeline merged \\
Organisation & Final artifact & code, tests, docs & Organisation--System & Propagated & Failure Impact, Stability & incident history and regression coverage after deployment \\
\bottomrule
\end{tabular}
\end{table*}


The sixth row is the case discussed in Section~2: the developer approves
Claude Code's architecture once and does not re-read it after Copilot has
implemented against it, yet remains a trustor with respect to that
architecture through the artifact chain. This relationship does not
appear as an edge in Figure~\ref{fig:chatdev-worked-example}, which
depicts direct handoffs only; the profile in Table~\ref{tab:profile}
makes it explicit.

\section{Future Directions}

The framework opens three directions for the SE community. Each addresses a layer at which current SE-agent practice may benefit from new artifacts: standards, protocols, and tools. Together they aim to make trust profiles a first-class object in agent development, deployment, and evaluation.

\textbf{D1: Trust handoff protocols for multi-agent pipelines.} A trust
handoff protocol may attach a metadata envelope to each inter-agent
artifact. A minimal envelope for the handoff between the CTO Agent and
the Programmer Agent in Section~2 might read:

\begin{quote}
\small
\ttfamily
\{ "artifact": "architecture-spec-v3", \\
\hphantom{\{} "producing\_agent": "claude-code", \\
\hphantom{\{} "configuration": "propagated", \\
\hphantom{\{} "properties\_claimed": ["contextual\_groundedness"], \\
\hphantom{\{} "evidence": "validated against repo schema", \\
\hphantom{\{} "unresolved": ["dependency X not version-checked"] \}
\end{quote}

The producing agent declares what configuration it believes it operated
under, what property evidence it can offer, and what it could not
verify; the consuming agent inherits this record rather than treating
the artifact as a bare input. This resembles existing agent-to-agent
protocols that carry identity and reputation
metadata~\cite{interagent2025trust}, but none of these currently carry a
configuration or property-evidence field; the envelope above extends
that substrate rather than replacing it. Open questions include where
the envelope should be validated, how a chain of envelopes should be
composed across more than two agents, and empirical study of failure
modes when the protocol is absent.

\textbf{D2: Configuration-specific trust evaluation.}

A configuration-stratified reading of an existing benchmark illustrates
the gap directly. SWE-bench reports whether an agent resolves an issue
once; it does not report whether the same agent, run three times on the
same issue, produces the same patch (Stability), or whether a passing
patch depends on an assumption about the repository that only surfaces
on a later, unrelated issue (Failure Impact under Propagated
reliance). A configuration-stratified suite would re-run a subset of
tasks under identical conditions to measure Stability directly, and
would track whether accepted patches are later reverted or amended by
downstream changes, as a proxy for Failure Impact once the patch has
propagated. Neither addition requires new tasks, only a different
reading of the same execution traces benchmarks already produce.
Calibration data from longitudinal developer
studies~\cite{liang2024icse,sabouri2025icse} may inform how such a
suite should weight the two measures.

\textbf{D3: Trust profile declarations.}

Machine-readable governance
for agents already exists in adjacent forms: Policy Cards encode
runtime constraints and crosswalk them to assurance
standards~\cite{mavracic2025policycards}, and agent registries such as
AgentHub bind capability claims to auditable evidence \cite{agenthub2025registry}. Neither
currently declares which \emph{trust profile} an agent is fit for ---
that is, which scenario--configuration pairs it has evidence for and
which it does not. An agent might carry strong evidence for Functional
Adequacy under Assistive use and no evidence at all for Failure Impact
under Propagated use; existing manifests have no field for this
distinction. A trust profile declaration would extend this substrate
with exactly that field: for each configuration an agent is intended to
operate under, which properties it claims evidence for, and the
strength of that evidence. This would let procurement, IDE integration,
and governance review consume a single artifact when deciding not just
whether to trust an agent, but where. Open questions include the
declaration schema, evidence-level definitions for each property, and
how such declarations would compose across a pipeline where several
agents' declarations must be read together, as in the workflow of
Section~2. A further use is role assignment: if declarations show which
properties different agents hold reliably, for example one agent strong
on Verifiability and another on Stability, they could inform which
agent occupies which role in a pipeline, rather than assigning roles by
convenience or availability alone.
\label{sec:researchagenda}

\section{Discussion}

The framework is conceptual. The three scenarios and four
configurations are closed by construction, as Section~4 states; the
five properties are not, and remain a starting point for further
refinement rather than a final taxonomy. Empirical work is needed to
validate whether this decomposition is sufficient and whether the
properties are correctly weighted across different deployments. No empirical standard applies to this paper, because it makes no
empirical claim: it proposes a structural decomposition and
instantiates it on an illustrative workflow, rather than testing a
hypothesis against collected data. The empirical work this framework
motivates, validating the property set, measuring configuration-stratified
benchmarks as sketched in Section~\ref{sec:researchagenda}, and studying
how the framework's profiles compare to developers' actual trust
judgements, belongs to follow-on studies with their own
instrumentation, not to this one.

\subsection{Property Dependencies}

The five properties are not independent, and the dependencies we have
identified run in a consistent direction: from properties that describe
the artifact itself toward properties that describe what happens when
the artifact is wrong. These are the dependencies visible from the
framework's own derivation; the list below is not claimed to be
complete, and other directional relationships among the five properties
may exist.

\textit{Contextual Groundedness constrains Functional Adequacy.} An
agent that misunderstands the project's APIs or conventions can produce
output that is functionally adequate in isolation but incorrect once
placed in context; grounding is therefore a precondition for adequacy
holding in practice, not a separate concern from it.

\textit{Verifiability constrains Failure Impact.} A failure that cannot
be traced to the decision that caused it is harder to detect and to
recover from; weak verifiability does not cause failures, but it
determines how expensive a failure turns out to be once one occurs.

\textit{Stability constrains the value of Verifiability.} Checking a
single run tells a practitioner little if the agent behaves differently
on the next run under the same conditions; verification is only
informative to the extent that what was verified is representative of
what will recur.

None of these dependencies are captured by treating the five properties
as a flat, unordered list. A practitioner who confirms Functional
Adequacy without first checking Contextual Groundedness, for instance,
may be confirming a property that will not hold once the artifact is
deployed.

\subsection{Structural Trustworthiness and Human Trust Perception}

Finally, the framework reasons about structural trustworthiness rather
than human trust perception, which varies with experience,
organisational culture, and interface design. Connecting the two is an
open problem that empirical studies of developer--agent interaction are
beginning to address~\cite{liang2024icse,sabouri2025icse}.

\section{Conclusion}

The rise of AI software engineering agents brings the classic
problem of trusting trust back at a new scale. In traditional software systems, trust may depend on compilers, libraries, and toolchains. In agentic software engineering, trust may additionally depend on models, tools, prompts, agents, generated artifacts, and multi-agent pipelines.

This paper proposed a structural framework for reasoning about trust in AI software engineering agents. By decomposing trust into three questions: \textbf{Why trust?},  \textbf{What trust?}, and  \textbf{How much trust?}, the framework helps distinguish between different deployment contexts and their corresponding trust requirements. The resulting model consists of three scenarios, four configurations, and five trust-relevant properties. The central claim is not that trust can already be fully measured, but that trust should first be structurally understood. A developer using autocomplete, an organisation deploying an autonomous repair agent, and a multi-agent pipeline coordinating software development do not face the same trust problem. The framework provides a way to identify the relevant trust relationship,
determine how reliance is configured, and select the properties and evidence
needed for context-aware trust evaluation in SE-agent deployments. Structured reasoning may therefore be a necessary first step towards measurable, context-aware trust evaluation for AI agents in software engineering.

\paragraph{AI Disclosure:} The ideas and contributions in this
paper are original. AI-based tools were used for refinement.

\bibliographystyle{ACM-Reference-Format}
\balance
\bibliography{ref}

\end{document}